# A Molecular Dynamics Investigation of Mechanical Properties of Graphene Reinforced Iron Composite and The Effect of Vacancy Defect Distance from the Matrix-Fiber Interface.

Raashiq Ishraaq[1,a)], Mahmudur Rashid[1,b)], A. M. Afsar[1,c)]

[1]Department of Mechanical Engineering, Bangladesh University of Engineering and Technology, Dhaka-1000, Bangladesh

Corresponding author: ishraaqraashiq@gmail.com
Corresponding author: mahmud.me.buet@gmail.com
Corresponding author: mdafsarali@me.buet.ac.bd

**Abstract.** Graphene is a material of excellent mechanical properties, which make it an ideal fiber for reinforcing metal. Since iron is the most used metal in the world, reinforcing iron with graphene can reduce the overall requirement of material in any application where strength is demanded. However, the effect of graphene reinforcement on the mechanical properties of iron needs to be known before the industrial application of the composite. In this paper, we have investigated the mechanical properties of graphene-reinforced iron composite by Molecular Dynamics (MD) method for various conditions. The properties were investigated by applying uniaxial tension on a modeled representative volume element (RVE). The findings in our study show that for only 9% fiber volume, the failure strength of iron increased 48.87% and Young's modulus 41.315%. The effect of temperature on the mechanical property of the composite was also studied because the knowledge is required for manufacturing products with the composite operating at a wide temperature range. MD analysis also revealed that the initiation of fracture is from the matrix-fiber interface. We also investigated how the distance of vacancy defects from the matrix-fiber interface affects the mechanical properties of the composite, which can be used to select a suitable manufacturing process. The results obtained from this study show that vacancy defects lower the strength at a greater extent as it gets closer to the interface.

## INTRODUCTION

Graphene is one of the strongest materials in the world due to its exceptionally high tensile strength and Young's modulus [1]. However, it is also very light in weight having a density around 1.06 gcm$^{-3}$. These properties of graphene make it a good reinforcement material. Previously, researchers mostly focused on graphene reinforced polymer composites. Recently graphene reinforced metal matrix composites are gaining attention due to the demand for materials having high strength and low weight in aircraft and automobile industries. Researchers conducted many studies on graphene reinforced metal matrix composites (GRMMC) having matrix metal of Al, Cu, Mg, Ni [2].

Molecular Dynamics (MD) is an efficient method to study the reinforcement effect of graphene. Recently, Rezaei [3] studied the mechanical properties of graphene-reinforced copper using MD. They found that the strength and Young's modulus of copper increased by 86% and 104.6% respectively. Duan et al. [4] also used MD method to find the effect of graphene layer number and chirality on the mechanical properties of graphene-reinforced copper. Mokhalingam et al. [5] studied the mechanical properties of graphene-reinforced aluminum by MD.

Though iron is the most used material in the world there are very few studies, which consider iron as the matrix material. Lin et al. [6] synthesized graphene oxide (GO) reinforced iron by laser sintering. They found that the surface micro hardness of iron increased by 93.5% for adding only 2 wt. % GO. Wang et al. [7] studied the effect of graphene layer number on the hardness of graphene reinforced iron using MD and reported a 139.5% increase in hardness for monolayer graphene. Wang et al. [8] also studied the interaction of edge dislocation with the graphene sheet in graphene-reinforced iron composite by MD simulations.

Considering the scarcity of studies regarding iron matrix, in this paper we investigated the mechanical properties (ultimate strength, Young's modulus, and failure strain) of graphene-reinforced iron composite under uniaxial tension using MD. At first, the increase of strength and stiffness due to graphene reinforcement was measured. In the automotive and aerospace industries the knowledge of a material's mechanical properties in different temperatures is required. Therefore, the variation of mechanical parameters with respect to temperatures was also measured. We also investigated the failure process of the composite. Vacancy defect is a common phenomenon while manufacturing composites. It can occur at different distances from the matrix-fiber interface depending on the manufacturing process. However, studies considering the effect of vacancy defect distance from the fiber-matrix interface are rare. Here we also investigated how the distance between the matrix-fiber interface and the vacancy defects present in the matrix affects the mechanical parameters of the composite.

## METHODOLOGY

A representative volume element (RVE) model of graphene reinforced iron composites was created which is shown in Fig.1. A single-layered graphene sheet was embedded in a rectangular iron block to model the RVE. Large-scale Atomic/Molecular Massively Parallel Simulator (LAMMPS) was used to create the model and perform simulations. OVITO was used for visualization. The dimension of the model is 54.27 Å, 114.75 Å and 57.12 Å in the X, Y and Z direction respectively. The RVE model contained 9% fiber volume.

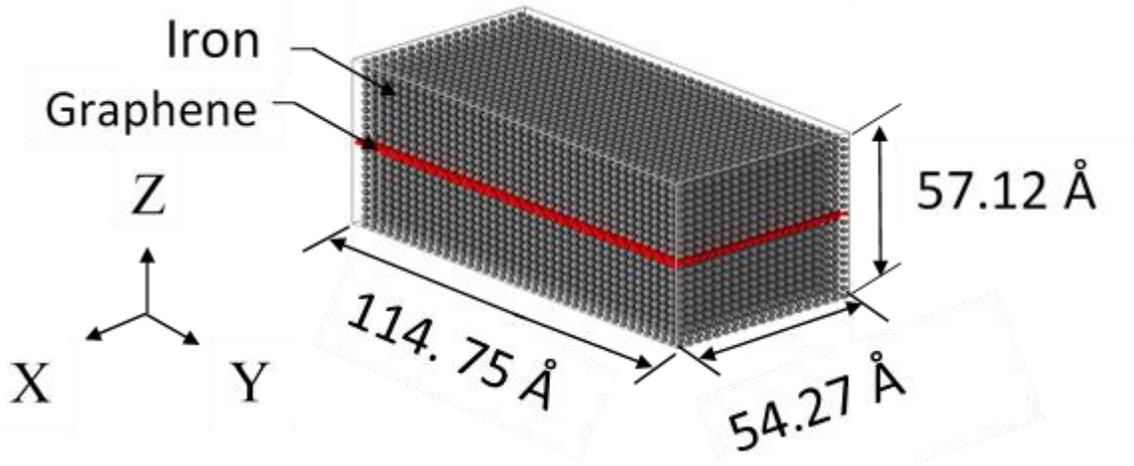

**FIGURE 1.** RVE model of graphene reinforced iron composite containing 9% fiber in volume.

To model the three different atomic interactions, three potentials were used. For Fe-Fe interactions, EAM (Embedded Atom Method) potential was used, which is described below.

$$E_i = F_\alpha \left( \sum_{j \neq i} \rho_\beta(r_{ij}) \right) + \frac{1}{2} \sum_{j \neq i} \emptyset_{\alpha\beta}(r_{ij}) \tag{1}$$

Here $F$ denotes the embedding energy, which is a function of electron density $\rho$. $\emptyset$ Is pair potential function and $r_{ij}$ is the distance between atom i and j. Subscripts α and β are element types of atom i and j. The EAM potential developed by Ackland et al. [9] was used for this study.

AIREBO potential was used for the C-C bonds in the graphene sheet. It consists of REBO, Lennard-Jones and torsional interaction potential as shown in equation 2.

$$E= \sum_i \sum_{i \neq j} [E_{ij}^{REBO} + E_{ij}^{LJ} + \sum_{k \neq i} \sum_{l \neq i,j,k} E_{i,j,k,l}^{TORSION}] \quad (2)$$

$E_{ij}^{REBO}, E_{ij}^{LJ}, E_{i,j,k,l}^{TORSION}$ are REBO, Lennard-Jones and torsional potential accordingly. The REBO potential describes the covalent bonding. The Lennard-Jones and torsional terms describe the intermolecular interaction and dihedral angle effect.

Most of the researchers consider Van-der-walls interaction to model interfacial bonding between matrix metal and graphene sheet. To model the Van-der-walls interaction previous studies used 6-12 Lennard-Jones potential [4,5,7,8]. So, for modeling Fe-C interaction at the interface 6-12 Lennard-Jones potential as described below.

$$E= 4\epsilon \left[ \left(\frac{\sigma}{r}\right)^{12} - \left(\frac{\sigma}{r}\right)^{6} \right] \quad (3)$$

$\epsilon$ is the potential well depth, $r$ is the interatomic distance and $\sigma$ is the distance where potential energy is zero. The values for $\sigma$ and $\epsilon$ are 2.221Å and 0.043 eV. These values were also used to model Fe-C interfacial interaction by Wang et al [7].

The radius of the hole inside the matrix was such that the graphene sheet and iron atoms maintain a distance of h=0.8584$\sigma$ as suggested by Jiang et al. [10] used by Mokhalingam et al. [5].

The total potential energy of the Fe-CNT system is

$$E_{TOTAL} = E_{CNT(airebo)} + E_{Fe(EAM)} + E_{INTERFACE(Lennard-Jones)} \quad (4)$$

After creating the model, energy minimization using the conjugate gradient method was done to ensure there is no overlap of atoms. A sequence of NVT and NPT equilibration was also performed for 100000 time steps each. The value of each time step was 1 fs. The value of the number of time steps ensured stable fluctuation of temperature and pressure. During the NPT equilibration, the temperature was held at 300K and pressure at 1 bar to simulate atmospheric conditions. Uniaxial tension was applied in the y direction at a constant strain rate. Strain rate of 0.001 ps$^{-1}$ was used for all the tensile test simulations in this study.

The mechanical stress of the iron-graphene composite system was calculated using the virial stress as described below

$$\sigma = \frac{1}{\Omega} \sum_i [-m_i \dot{u}_i \otimes \dot{u}_i + \frac{1}{2} \sum_{i \neq j} r_{ij} \otimes f_{ij}] \quad (5)$$

Here m is the mass of the atom i and $\dot{u}_i$ is the time derivative of the displacement vector $u_i$. $f_{ij}$ is the interatomic force on atom i by atom j and $r_{ij}$ is the distance between them.

## VALIDATION

To validate the procedure in this study uniaxial tension test was done on a graphene sheet with 54.27 Å in the x and 114.75 Å y direction. The mechanical properties were compared with the available literature. A similar method for validation was used by Mokhalingam et al. [5] when validating their study on graphene reinforced aluminum composite. Periodic boundary condition was applied in the x and y dimension. The results obtained from this simulation and from an experimental study of graphene by Lee et al [1] are shown in table 3. The results show errors within 5 %, which is quite negligible.

**TABLE 1.** Comparison of mechanical parameters of graphene from the current study and different literature.

| Mechanical parameters | Results obtained in this study (GPa) | Lee et al (GPa) |
|---|---|---|
| Elastic modulus | 964.759 | 1000 |
| Ultimate stress | 114.7 | 130 ±10 |

# RESULTS AND DISCUSSION

## Comparison of Mechanical Properties and Effect of Temperature:

To measure the enhancement of mechanical properties because of graphene reinforcement, the RVE model of the composite and a pure iron block of similar size was compared. Fig. 2 shows the stress-strain relationship of pure iron and graphene reinforced iron composite. The failure of any component (matrix or fiber) was considered as the failure of the composite. In this case, the matrix (iron) fails first. Therefore, the stress developed in the composite when the matrix fails was considered the ultimate stress. In Fig. 2 stress-strain relationship of the composite was plotted until the failure of the matrix.

The stress-strain relation from Fig. 2 shows an exceptional increase in strength and stiffness for adding only 9 % fiber volume. The results show that the ultimate strength increases 48.87 %, the Young's modulus increases 41.315 % but the ultimate strain decreases by 8.31 %. The possible reason for the decrease in ultimate strain is the weakening of the matrix near the matrix-fiber interface, which causes the slip planes to activate at an earlier strain.

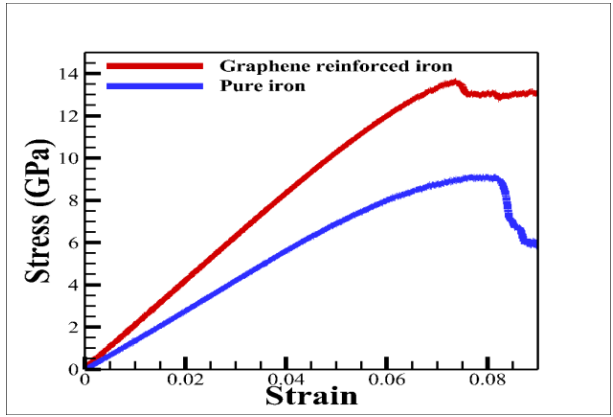 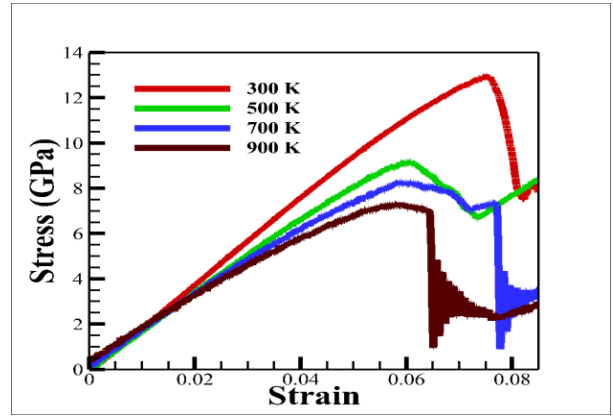

**FIGURE 2.** stress-strain relation of pure iron and the composite containing 9% fiber volume percentage.

**FIGURE 3.** stress-strain relation of the composite at various temperatures

To investigate the effect of temperature on the mechanical properties, uniaxial tension was applied on the RVE model at four different temperatures (300K, 500K, 700K, and 900 K). Figure 3 shows the stress-strain relationship of the composite at different temperatures. At high temperatures, the atomic vibration increases that create more voids and defects in the matrix. As a result, failure initiates at lower stress that causes the ultimate strength and stiffness of the composite to fall rapidly. At higher temperatures, strength of graphene also decreases due to weakening of the atomic bond because of increased atomic velocity. For every 200K increment of temperature, the ultimate strength decreases on an average of 16.61 % and Young's modulus decreases 8.44 % than the previous temperature. However, from Fig.3 3 it can be seen that at higher temperatures more plasticity occurs during the failure.

## Failure Mechanism of the Composite:

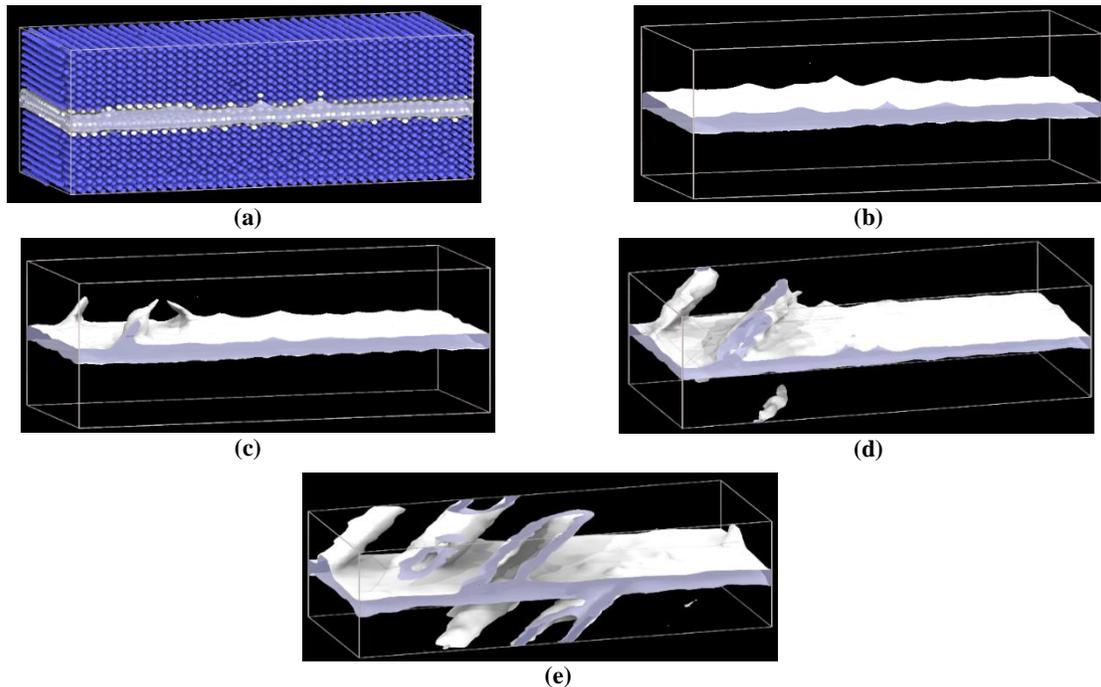

**FIGURE 4.** (a) shows DXA analysis on the unstrained RVE where white mesh covering particles that deviates from ideal bcc configuration. (b) shows the white mesh without atoms. (c) shows the initiation of failure in the matrix at 7% strain. (d) Represents the propagation of the failure. (e) shows the initiation and propagation of dislocation from multiple failure sites.

In this composite, the iron fails first due to its lower failure strain than graphene. For investigating the origin and nature of failure, dislocation extraction algorithm (DXA) modifier of OVITO software was used. In the DXA modifier, parts of the crystal that deviates from ideal bcc configuration are covered in a white mesh.

Figure 4(a) shows the condition of the RVE model before applying uniaxial tension. Since the graphene and the interface atoms deviate from ideal bcc configuration, it is covered in white mesh. To observe the initiation and propagation of failure, the atoms were hidden and only the white mesh is made visible from Figs 4(b) to 4(e). Figure 4(c) shows the initiation of dislocation propagation at 7% strain. It also shows that the defect originates from the interface. Choi et al. [11] observed similar phenomena for Al-CNT composite. After the initiation of the defect, it propagates in 45-degree plane that is also the maximum shear stress plane as shown in Fig. 4(d). At 7.2% strain dislocations originate from other parts of the interface and propagate in a similar manner, which can be seen in Fig. 4(e).

## Effect of Distance Between the Matrix-Fiber Interface and Vacancy Defects:

The RVE is symmetric about the graphene sheet. Therefore, any analysis on the matrix portion of either side of the graphene sheet will give similar results provided all the conditions remain the same. Considering this fact, the analysis was done only on one side of the matrix, which is below the graphene sheet.

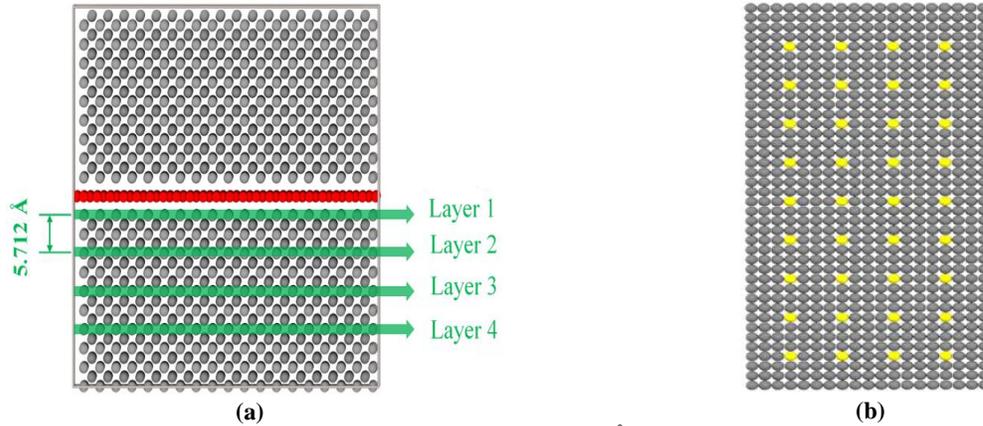

**FIGURE 5.** (a) shows the four layers of atoms separated by 5.712 Å. The position of vacancy defects (yellow) in a layer is shown in (b).

Four layers of atoms were selected at different distances from the matrix-fiber interface with a separation of 5.7 Å as shown in Fig. 5 (a). Four uniaxial tension simulation was performed on the RVE model. During each simulation, only one out of the four layers of atoms was selected and 36 atoms were deleted from the selected layer. Each time a different layer was selected which ensured that the vacancy defects were introduced in the matrix at a different distance from the matrix fiber interface.

Figure 6 (a) shows the stress-strain relation obtained from the simulations and Fig. 6 (b) shows how the ultimate strength and Young's modulus vary when the vacancy defects are in different distances from the matrix-fiber interface. From Fig. 6(b) it can be seen that Young's modulus increases drastically when the vacancies move from first atomic layer to the second and then becomes almost constant.

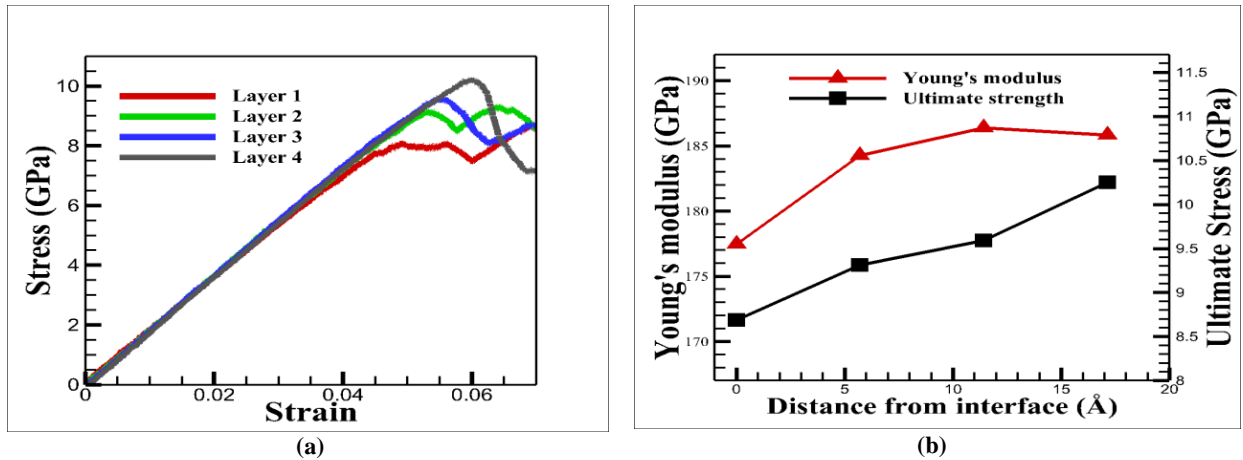

**FIGURE 6.** (a) shows the stress-strain relation of the composite for having vacancy defects at different distances from the interface. (b) shows the variation of Young's modulus and ultimate stress with respect to the distance between the defects and matrix-fiber interface.

However, the ultimate strength keeps increasing noticeably, as the vacancy defect gets further from the interface. This phenomenon can be explained due to the weakness of the matrix near the interface. The atoms of the matrix near the interface are already deviated from ideal bcc configuration, which causes the atoms to move from their lattice position easily and activate slip systems. Therefore, when more defects are present near the interface it further weakens the matrix at that position causing it to fail at much lower stress.

# CONCLUSION

In this study, the mechanical property of graphene-reinforced iron composite was investigated by MD analysis. From the simulations, it is evident that strength and stiffness increase drastically for reinforcing iron with graphene. At higher temperatures, the composite exhibited more plasticity near the failure stress. Analysis of the failure process also revealed that the failure initiates from the matrix-fiber interface. A detailed investigation also exhibited that the vacancy defect lowers the strength and stiffness of the composite at a noticeable rate as it gets closer to the composite. This RVW model of iron-reinforced graphene composite can be used also to measure the effect of matrix volume fraction, fiber orientation, fiber-to-fiber distance and defects in the graphene.